\documentclass[12pt]{article}

\usepackage{amsmath,amssymb,exscale}
\usepackage{array,multicol}
\usepackage{afterpage,float,flafter}
\usepackage{epsfig,rotating,pifont}
\usepackage{cite}
\def\beq{\begin{eqnarray}}
\def\eeq{\end{eqnarray}}

\def\e{{\epsilon}}

\def\lsim{\mathrel{\rlap{\lower3pt\hbox{\hskip0pt$\sim$}}
    \raise1pt\hbox{$<$}}}         
\def\gsim{\mathrel{\rlap{\lower4pt\hbox{\hskip1pt$\sim$}}
    \raise1pt\hbox{$>$}}}         

\def\e{{\rm e}}

\newcommand{\be}{\begin{equation}}
\newcommand{\ee}{\end{equation}}
\newcommand{\bea}{\begin{eqnarray}}
\newcommand{\eea}{\end{eqnarray}}
\newcommand{\bg}{\begin{gather}}
\newcommand{\eg}{\end{gather}}
\newcommand{\bseq}{\begin{subequations}}
\newcommand{\eseq}{\end{subequations}}

\renewcommand{\ln}{\mathop{\rm ln}\nolimits}

\title{
{\LARGE\textbf{Gravity Cutoff 
in Theories\\ with Large Discrete Symmetries} }
\author{
\textbf{Gia Dvali,{\boldmath $^{a,b}$} 
 Michele Redi,{\boldmath $^c$} Sergey
  Sibiryakov{\boldmath $^{\,a,d}$}}\\ 
 \textbf{and Arkady Vainshtein{\boldmath $^{a,e}$}}\\  
$^a$\normalsize\emph{Theory Group, Physics Department, CERN, CH-1211 Geneva 23,
  Switzerland}\\
$^b$\normalsize\emph{Center for Cosmology and Particle Physics, 
Department of Physics,}\\
\normalsize\emph{
New York University, New York NY 10003, USA}\\
$^c$\normalsize\emph{Institut de Th\'eorie des Ph\'enomenes Physiques,
EPFL, CH-1015, Lausanne, Switzerland}\\
$^d$\normalsize\emph{Institute for Nuclear Research of the Russian
  Academy of Sciences,}\\ 
\normalsize\emph{60th October Anniversary prospect, 7a, 117312 Moscow,
  Russia}\\
$^e$\normalsize\emph{Fine Theoretical Physics Institute, School of Physics and Astronomy}\\ 
\normalsize\emph{University of Minnesota, 116  Church St SE, Minneapolis, MN 55455, USA}
}}

\date{}
\begin{document}
\maketitle \thispagestyle{empty}

\vspace{-14cm}

\begin{flushright}
CERN-PH-TH/2008-065\\
FTPI-MINN-08/11\\ 
UMN-TH-2642/08\\
\end{flushright}

\vspace{12cm}

\begin{abstract}
We set an upper bound on the gravitational cutoff in theories with 
exact quantum numbers of large $N$ periodicity, such
as $Z_N$ discrete symmetries. The bound stems from black hole
physics. It is similar to the bound appearing in theories with $N$
particle species, though a priori, a large discrete symmetry does not
imply a large number of species.  
Thus, there emerges a potentially wide class of new theories, 
that address the hierarchy problem by lowering the gravitational
cutoff due to  existence of  
large $Z_{10^{32}}$-type symmetries.

\end{abstract} 

\newpage
\renewcommand{\thepage}{\arabic{page}}
\setcounter{page}{1}

\section{Introduction}

Black hole (BH) physics is a powerful tool for extracting
non-perturbative information about microscopic structure of the
theory. As examples of such use of BHs one may list the argument
about violation of continuous global symmetries in gravitational theories
\cite{nohair}, the bound on the entropy of bounded systems
\cite{Bekenstein:1980jp}, and the constraints on possible violation of
Lorentz invariance \cite{Dubovsky:2006vk,Eling:2007qd}. Yet another example 
\cite{bound} is the restriction on the number $N$ and
mass $M$ of particle species, which in the case 
of stable species and large 
$N$ reads  
\begin{equation}
\label{PM}
N M^2 \lesssim M_P^2\;,
\end{equation}
up to a factor that scales as $\sim$ $\ln N$. Here
$M_P$ is the Planck mass. 
 
As also shown in \cite{bound},  the same BH bound applies even to a
single species of 
mass $M$ that carries an exactly conserved quantum 
number (not associated with any long-range classical gauge force)  of
periodicity $N$. An example of such a quantum number  
can be a discrete gauge symmetry $Z_N$\cite{ZN1,ZN3}, 
or a quantum hair
under some massive integer spin field\cite{quantum}.  
In what follows,  we shall investigate this situation. 
Thus, unless otherwise stated, $N$ will denote periodicity of $Z_N$
(or of some other  
exact quantum number) and not the number of particle species 
and the bound (\ref{PM}) should be understood accordingly.    
 
The above bound is applicable, in particular, to scalar fields and 
implies that masses of $N$ scalar fields, or of a single scalar
field charged under a $Z_N$-symmetry, are automatically limited by
$M_P/\sqrt{N}$. 
By naturalness arguments presence of light scalar fields suggests
existence of some new stabilizing physics at that scale. 
For the case of many species this statement was made
rigorous in \cite{dr} where it was shown that in a theory with 
large number of
species the gravitational cutoff comes down to $M_P/\sqrt{N}$.
(This is consistent with earlier perturbative arguments 
\cite{Dvali:2000xg,Veneziano:2001ah}.)
The purpose of this note is to prove a similar statement for the
case of large discrete symmetry.

\section{Black Hole  Argument} 
\label{Sec:2}

Consider a (scalar) field $\phi$ of mass $M$ transforming under a
discrete  symmetry  $Z_N$,   
\begin{equation}
\label{zn}
\phi \, \mapsto \, \phi \,  {\rm e}^{\,i\,\frac{2\pi}{N}\,k}\,\qquad k=0,\ldots, N-1\,.
\end{equation}   
We assume that this $Z_N$ symmetry is exact, i.e. it is not 
 violated at any scale.  
A straightforward way to ensure this 
is to declare that $Z_N$ is a gauge symmetry. 
However,  for our reasoning it is unimportant what underlying physics
guarantees exactness of $Z_N$.  
We then make the two following assumptions: \\
{\it a}) the particle $\phi$ has the largest charge to mass ratio among all the
particles carrying $Z_N$ charge;\\ 
{\it b}) there are no BH remnants.
    
We are going to prove that there is a bound on the cutoff $\Lambda$ of
the low energy theory,   
\begin{equation}
\label{newbound}
\Lambda \leq \frac{M_P^2}{NM}\;.
\end{equation}
If $M\sim M_P/\sqrt{N}$ this bound coincides with the 
bound
\be
\label{bound1}
\Lambda\leq \frac{M_P}{\sqrt N}
\ee 
implied by naturalness. However, in general, the bound (\ref{newbound}) 
is weaker than (\ref{bound1}). 
We show that the stronger bound (\ref{bound1}) is obtained
if one makes an additional assumption that the property of 
negative heat
capacity of BHs persists in the high-energy theory.

Let us proceed to the proof of Eq.\,(\ref{newbound}). 
One performs the same thought
experiment \cite{bound} as in establishing the bound (\ref{PM}).  
Take a
macroscopic (arbitrarily large) BH and throw a number $\sim N$ of
$\phi$-particles into it. In this way we endow the BH with
$Z_N$-charge of order $N$. Then one waits for the BH in question to
evaporate. Since the $Z_N$-symmetry is exact at all
scales and there are no remnants the BH eventually has to return the
exact amount of the swallowed charge. Indeed, if the
returned charge were not equal to the original one the BH would
mediate a process that violates $Z_N$ explicitly, in contradiction
with our assumption.
  
The crucial point is that as long as the BH is the usual
Schwartzschild BH (SBH) it cannot give out any $Z_N$-charge. Indeed,
radiation of SBH is thermal and contains as many $\phi$-particles as
the antiparticles. Thus, to return back the $Z_N$-charge, the
properties of BH must get modified when it reaches a certain size,
$R_{BH}\sim\Lambda^{-1}$. This implies existence of a new
physics at
the scale $\Lambda$; in other words, $\Lambda$ is a cutoff of the
low-energy theory.

Consider the BH that has just reached the cutoff scale. The
mass of the BH at this moment must be sufficient to produce $\sim N$
of $\phi$-quanta,
\be
\label{MBH}
M_{BH}\geq NM\;.
\ee  
On the other hand, the BH mass and size are still related at this
moment by the standard Schwartzschild expression,
\be
\label{RBH}
M_{BH}\sim R_{BH}M_P^2\;.
\ee 
Combining Eqs.~(\ref{MBH}), (\ref{RBH}) one obtains the bound
(\ref{newbound}). 

Notice that the above proof is UV-insensitive in the sense that it does
not depend on the precise nature of BHs that are smaller than the
cutoff scale $\Lambda^{-1}$. 
All we have used is the conservation of energy
which is entirely a large-distance constraint.

The bound can be improved if we make an additional assumption that the
property of negative heat capacity of BHs persists in the high-energy
theory. More precisely, we assume that
the BH, after it reaches the size $R_{BH}\sim\Lambda^{-1}$
corresponding to the Hawking temperature $T_H\sim\Lambda$, continues
to radiate preferentially into modes with energies equal or higher than
$\Lambda$. Then, Eq.\,(\ref{MBH}) is replaced by 
\be
\label{MBH1}
M_{BH}\geq N\Lambda\;.
\ee
When combined with (\ref{RBH}) it yields the stronger bound
(\ref{bound1}). 

The above assumption about the BH spectrum 
appears to be natural. Its violation would imply
very unusual properties of small BHs: they should be very cold and decay
into quanta with inverse momenta greatly exceeding the size of the
BH. Although we cannot exclude such a possibility, we conclude that
under reasonable assumptions about the properties of small BHs the
bound on the cutoff scale is (\ref{bound1}). 

It is worth comparing the argument presented in this section with the
case of large number
of species \cite{dr}. In the latter case the existence of the low cutoff
can already be established
in perturbation theory by considering the graviton propagator.
The loop corrections to the propagator are amplified by the large number
of species
and the perturbative expansion goes out of control precisely at the scale
$M_P/\sqrt{N}$ signaling that the cutoff is reduced to this value.
The same conclusion can also be inferred directly from BH physics.
The Hawking radiation
of a BH of size $(M_P/\sqrt{N})^{-1}$ 
is drastically amplified as it can radiate $N$ species.
As a consequence such BH would have a lifetime of order of its size and
therefore it is not a classical
object as in ordinary general relativity. On the other hand, in the case
of a single field charged under a large discrete symmetry 
the perturbation theory does not show any sign of
breaking down. Similarly, there is no indication of the 
radiation of a BH of the size 
$(M_P/\sqrt{N})^{-1}$ blowing up.
Nevertheless as we have shown the
consistency of the theory requires
the presence of a low cutoff. The argument that enables to establish
the existence of the cutoff is intrinsically non-perturbative and uses
in an essential way the BH physics. This is reminiscent of 
the ``gravity as the weakest force'' conjecture \cite{nima}. It
would be interesting to explore a possible
connection between this conjecture and our work.

\section{Explicit Examples} 
 
In this section we consider a few examples of theories with large
discrete symmetries.

1) Consider a $U(1)$ gauge symmetry with two scalar fields, $\phi$ and
$\chi$. The field $\phi$ has a unit charge $e$ while the charge of the
field $\chi$ is $Ne$. Let the $\chi$-field develop a non-zero vacuum
expectation value (VEV), thus breaking the $U(1)$ symmetry down to
$Z_N$. The latter symmetry acts on $\phi$ according to
Eq.\,(\ref{zn}). The field $\phi$ is assumed to be 
lighter than the other fields, so it is the
only degree of freedom at low energies.

It is important to notice that setting the ratio of charges of the
fields $\chi$ and $\phi$ to a rational number (which we, for
simplicity, took to be an integer) is not a fine-tuning. Rather, this
is required by the bound (\ref{PM}). Indeed,    
if the ratio of charges were an irrational number
the effective discrete symmetry would be $Z_{\infty}$, which is
impossible. 

In this theory the $Z_N$-charge
inside a given volume of space can be monitored in the following way.  
Because of the non-trivial topological structure of the
vacuum manifold (non-contractible  loops) there are cosmic strings in this
theory, around which the phase of the $\chi$ VEV winds by $2\pi$
multiple. These cosmic strings contain a unit flux of the gauge field. 
This  allows to monitor the $Z_N$-charge of a system
through the Aharonov-Bohm (AB) effect in the scattering of the cosmic
strings from the system\cite{ZN1,ZN3}. 
In particular, if the system collapses into a BH, the latter has a
quantum $Z_N$ hair \cite{ZN1,ZN3}
that the AB effect can probe.

The proof of Sec.~\ref{Sec:2} implies that gravity in this model
must be modified at distances $(M_P/\sqrt{N})^{-1}$. Indeed,
from the
proof
it is clear that BHs with the size smaller than 
$(M_P/\sqrt N)^{-1}$
have to acquire hair capable of producing $Z_N$-charge asymmetry in
the BH evaporation. On the other hand, such hair are impossible in 
Einstein's general relativity. At the classical level this follows
from the no-hair theorems \cite{nohair}. Quantum effects do not help
either.  Indeed, existence of quantum hair leads to polarization of
vacuum around 
BHs with $Z_N$-charge
\cite{ZN3}. This vacuum polarization is sensitive to the $Z_N$-charge
of the BH and, a priori, can contribute to the asymmetry of the
evaporation. However, at weak coupling, $Ne\ll 1$, 
the effect is exponentially suppressed \cite{ZN3}
and is unable to produce the
necessary asymmetry. Thus we conclude that the physics responsible for
the cutoff at $M_P/\sqrt{N}$ must involve gravity in an essential way.

2) Existence of a large discrete symmetry may be accompanied by the
presence of a large number of species in the theory. Then, the latter
property, by itself, implies a low gravitational cutoff
\cite{dr}. This point is illustrated by the following example.            
  
Consider an $SU(2)$ gauge theory with two scalar fields, $\phi_j$
and $\chi_{j_1j_2...j_N}$, transforming as a fundamental and $N$-rank
symmetric tensors respectively. Here $j=1,2$ and $j_k = 1,2,~~ k =1...N$
are fundamental indices. We assume that the field $\chi$
acquires a VEV of only one component $\chi_{11...1}$. 
This VEV breaks the continuous $SU(2)$-symmetry down to a
discrete $Z_N$ factor, under which 
$\phi_1 \mapsto\phi_1\, \e^{i\frac{2\pi}{N}}$ 
and $\phi_2 \mapsto\phi_2\, \e^{-i\frac{2\pi}{N}}$.

One may be tempted to apply our argument to show that the
gravitational cutoff in this theory is low using the field
$\phi_1$ (or $\phi_2$) in the proof. However, it would be
incorrect: the proof of Sec.~\ref{Sec:2} is not
directly applicable to this case. The reason is that the theory
contains particles with arbitrarily large $Z_N$-charges and so
the assumption {\it a}) of the proof is violated. 
The states with large $Z_N$-charges are the components of the
field $\chi$.
Indeed, a component $\chi_{j_1j_2...j_N}$ with $n$ indices equal to 
$1$ and remaining $N-n$ indices equal to $2$ carry $2n-N$ 
units of the $Z_N$ charge. Correspondingly, a discrete charge of 
arbitrary $2n-N < N$ number of the $\phi_1$ fields can be recycled 
by a BH into a single $\chi$ quantum. The corresponding gauge
invariant operator has the form 
\begin{equation} 
\bar\phi^{j_1}...\bar\phi^{j_{2n}}  
\chi_{j_1...j_n a_1...a_{N-n}}\chi_{j_{n+1}...j_{2n} b_1...b_{N-n}}
\epsilon^{a_1b_1}...\epsilon^{a_{N-n}b_{N-n}}\;.
\end{equation}

However, the gravitational cutoff is still lowered down to $M_P/\sqrt N$ in
this model. This is due to the fact that the theory contains $N$
species which are the $N$ components of the symmetric tensor $\chi$. 
Thus we again find in this example that large discrete symmetry
implies cutoff $M_P/\sqrt{N}$, though, in this case, indirectly: through a
large number of species. 
     
3) The previous example suggests that the condition {\it a}) of the
proof can be
replaced by a weaker one. Going through the proof
it is straightforward to convince oneself that the following
requirement for the structure of the theory 
is sufficient: it should be impossible to reproduce an
arbitrary $Z_N$-charge with a small (much less than $N$) number of
particles belonging to a small number of species. The following
example proposed in \cite{bound} demonstrates that this condition
cannot be weakened further.

Consider a theory with $n$ scalar fields $\Phi_k$, $k=1,\ldots,n$,
with the following sequence of couplings,
\begin{equation}
\label{sec}
\Phi_1^3 \, + \, \Phi_1^* \Phi_2^3 \, + \, \Phi_2^* \Phi_3^3 \, + \, ... \,  \Phi_k^*\Phi_{k\, +\,1}^3 \, +  ...
\, + \, \Phi_{n\, -\,1}^*\Phi_n^3\,+\, h.c. \, .
\end{equation}
This theory is invariant under $Z_N$ symmetry with $N=3^n$ and the
transformation law,
\begin{equation}
\label{zncharge}
\Phi_k \mapsto  \Phi_k\,\e^{i3^{n-k}\frac{2\pi}{N}} \, .
\end{equation}
In other words, the $Z_N$-charge of the field $\Phi_k$ is $3^{n-k}$.

Let $\Phi_n$ be lighter than the rest of the fields. Then, integrating
out the first $(n-1)$ fields one obtains the effective $Z_N$-invariant
theory for $\Phi_n$. However, in this case one cannot use the BH
argument to claim that the 
mass of $\Phi_n$ is bounded by $M_P/\sqrt{N}$, nor that there is a
gravitational cutoff at $M_P/\sqrt{N}$. Indeed, any $Z_N$-charge of order $N$
thrown into the BH in the form of $\Phi_n$ quanta can be radiated away
in the form of just of order $n\sim \ln N$ quanta of different
$\Phi_k$ fields at the late stage of BH evaporation. The arguments of
\cite{dr} are also not applicable because the total number of
species is only $\sim\ln N$. The best one can do in these
circumstances is to conclude that the cutoff of the theory does not
exceed $M_P/\sqrt{\ln N}$.    

4) The bound (\ref{bound1}) has been obtained without references to
the explicit structure of the theory of quantum gravity. Hence, by
consistency, it should be satisfied in the string theory. Here we
propose a simple example which shows that this is indeed the case.  
Consider the setup where
the $Z_N$ group is generated by an isometry of compact space in  
string theory compactification. We take the string coupling to be of
order one so that the 10-dimensional Planck mass is set by the string 
scale $M_S$. Consider now a compactification on $T_6 \times {\cal
  M}_4$, where $T_6$ is a 6-dimensional torus and ${\cal M}_4$ is the
4-dimensional Minkowski space. 
The isometry group of this space is $U(1)^6$. 
We wish now to break one of the $U(1)$'s down to $Z_N$. 
Let the radius of the corresponding circle be $R$. 
We assume the radii of the other tori to be of order the string length. 
Then the relation between the 4-dimensional
Planck mass and the string scale is 
\be 
\label{PS}
    M_P^2 = M_S^2\,  (RM_S)\;.
\ee 
Let us break the $U(1)$ isometry on the $R$-circle down to $Z_N$ 
by creating $N$ fixed points around the circle. Alternatively this can
be done 
by placing $N$ identical branes and requiring the exact symmetry under
cyclic shifts.  Since the distance between the fixed points
or the branes is bounded by the string scale, the
maximal number of them that can be fitted on the circle is $N \leq RM_S$.
Recalling that $M_S$ is the cutoff of the low-energy effective theory
one sees that (\ref{PS}) reproduces the bound (\ref{bound1}).

\section{Implication for the Hierarchy Problem} 

The results of this paper shed new light on the proposal \cite{bound}
to solve the 
hierarchy problem by postulating a
large discrete symmetry with $N \sim 10^{32}$.  We find that in this
case the gravitational cutoff of the theory is not far from the weak
scale, thus the latter is automatically stabilized. 
A generic prediction of this solution to the hierarchy problem is
appearance of strong gravitational physics not far from the weak
scale. From the experimental point of view this physics is expected to 
manifest itself
in softening of
the scattering amplitudes at energies above the scale $M_P/\sqrt{N}$.
We now briefly discuss some aspects of implementing the above idea in
model building.  

From the constructive point of view it is desirable to have an
explicit mechanism ensuring the hierarchy between the Planck and the
weak scales.
In the Standard Model (SM) the Higgs boson cannot
transform under any exact symmetry, 
so it is impossible to give the $Z_N$-charge to the Higgs itself to
apply the bound (\ref{PM}) directly to its mass. Thus the idea is to
ascribe the $Z_N$-charge to some other fields whose mass gets
contributions from the Higgs VEV. The bound (\ref{PM}) on the mass of
these particles then implies the bound on the Higgs VEV.

Let us stress that in pursuing this strategy 
one should be careful not to run
into conflict with the low gravitational cutoff. To illustrate what we
mean let us consider the following example. One can
identify large $Z_N$ with 
the subgroups of the existing
global symmetries of the SM that would appear exact in
the absence of gravity. 
Ignoring gravity, the SM has two
classically-exact continuous global symmetries that account for
baryon and lepton number conservations. Thus one possibility is to
declare that the $Z_N$ symmetry in question is a subgroup of
some combination of baryon and lepton number symmetries. It is 
most straightforward to embed $Z_N$ into the $B-L$ symmetry because
the latter is automatically anomaly free. 
Then, from Eq.\,(\ref{PM}) the bound on $N$ is $M_P^2/m_\nu^2\gtrsim 10^{54}$ 
where 
$m_\nu \lesssim$ eV
is the mass of the lightest neutrino. Postulating the $Z_N$ symmetry 
with $N\sim 10^{54}$ we would
prevent the Higgs VEV to be larger than $10 -100$ TeV
since large Higgs VEV would make neutrino 
heavier\footnote{Note that in this
  construction neutrinos must be of the Dirac type since we
  assume an exact conservation of a subgroup of the $B-L$ 
  symmetry.} 
than the BH
upper bound for $N \sim 10^{54}$. However, according to the results of
this paper, such a large $N$ would lower the gravitational cutoff below
the weak scale, in contradiction with the observations. If we
want the cutoff at an acceptable level, we have to chose 
$N\sim 10^{32}$. In this case the direct BH bound on the Higgs VEV is much
higher than the bound on the cutoff. 

The reader may be puzzled why one should worry about applying the BH bound
(\ref{PM}) directly to the Higgs VEV, given the fact that in the
above example with $N\sim 10^{32}$ the cutoff is at the needed level?
Seemingly, the latter would suffice to solve the hierarchy problem.  The
point is that, in general, the cutoff controls the radiative stability
of the weak scale but need not necessarily constrain its tree-level
value.  Correspondingly, if the tree-level value is large the
physical scale will also be large even though the radiative
corrections are small. Hence, the small cutoff does not necessarily
guarantee the smallness of the physically observable weak scale,
whereas the direct BH bound on the mass (\ref{PM}) does.

To make our reasoning more transparent it is useful to make a
parallel with a much more familiar example of the low energy
supersymmetry. The cutoff that controls the radiative corrections to
the Higgs mass is the supersymmetry breaking scale in the observable
sector, $m_{susy} \sim $ TeV.  However, smallness of this cutoff
cannot explain why there is no large tree-level contribution to the
Higgs mass.  The latter puzzle is the essence of the celebrated
$\mu$-problem.  Thus, in order to solve the hierarchy problem in
supersymmetry, smallness of $m_{susy}$ is not enough. One needs an
additional mechanism that would guarantee smallness of $\mu$.  In our
case the analog of $m_{susy}$ is the low gravity cutoff $M_P/\sqrt N$.
However, the physical weak scale is restricted by the BH bound on the
particle masses.  Whenever we can directly apply this bound to the
weak scale, the hierarchy problem is solved, with no need of any
further assumptions about the tree-level masses versus cutoff.

An example when there is no large discrepancy between the cutoff and
the direct BH bound on the Higgs VEV is obtained in the following
way. 
The idea is to extend the SM by introducing new particles
charged under $Z_N$ that get their masses from the Higgs
VEV. These particles can be either scalars or fermions. In the
latter case we can postulate the existence of a vector-like pair of
left-handed lepton-like doublets $L, L'$ with opposite
hypercharges and two pairs of right-handed singlets: ``neutrinos''
$\nu,\nu'$ and ``electrons'' $e, e'$.  The couplings allowed by the gauge
symmetry are similar to the ones of ordinary 
leptons\footnote{The bare mass terms can be forbidden by invariance
  under a parity transformation which changes the sign of the 
  primed fields.}
\begin{equation}
\label{couplingstohiggs}
H L \nu  \, +  \, H^*\epsilon L e \,  + \, H^*L'\nu' \, + \, H\epsilon L' e' . 
\end{equation}
The Yukawa couplings are assumed to be of order one. The only difference
between these new vector-like fermions and the ordinary leptons is
that they transform under an exact $Z_N$ symmetry
with $N \sim 10^{32}$. This symmetry prevents mixings of the new fermions 
with 
other generations, and also implies the BH bound on their mass which
translates into a bound on the Higgs VEV.  In other words the BH
bound on the particle masses guarantees the stable hierarchy between
the Higgs VEV and the Planck mass. 
 
An alternative, even simpler, possibility is to introduce a 
scalar $S$ transforming under $Z_N$-symmetry and having no charge
under the SM gauge group. This scalar gets a mass from the Higgs
VEV through the following coupling in the Lagrangian
\begin{equation}
\label{StoHiggs}
- \lambda H^*H S^*S  \, - \, M^2S^*S,  
\end{equation}
where $\lambda \sim 1$ is the coupling constant and $M^2$ is the bare mass.
Equation (\ref{PM}) yields a bound on the total mass of $S$, 
$$ \lambda H^*H +
M^2 \lsim M_P^2/N\;,$$ 
which for $\lambda > 0, M^2 > 0$ translates into
the bound on the Higgs VEV $ \langle H \rangle \lsim M_P/\sqrt{N}$.
The latter is of the same order as the bound on the cutoff.

To avoid confusion let us stress that the solution of the hierarchy
problem considered above, based on the existence of a discrete
symmetry $Z_N$ with large $N$, is physically different from the
explanation of the hierarchy considered in Refs.~\cite{dr,CP} where
the large number $N$ was the number of particle species. In
particular, the arguments presented in \cite{CP} to show that 
the many-species scenario can
simultaneously solve the strong CP problem, 
are not directly applicable to the case of
large discrete symmetry. It would be interesting to understand whether
an alternative argument exists which could explain the smallness of
the strong CP parameter in the large $Z_N$ case as well.

\subsection*{Acknowledgments}
We thank S.~Dubovsky and G.~Gabadadze 
for useful discussions and comments. 
The work is
supported in part by David and Lucile Packard Foundation Fellowship
for Science and Engineering, by NSF grant PHY-0245068, 
by EU 6th Framework Marie Curie
Research and Training network "UniverseNet" (MRTN-CT-2006-035863)
and by DOE grant DE-FG02-94ER408.


\end{document}